\documentclass[preprint,showpacs,preprintnumbers,amsmath,amssymb,superscriptaddress,prb]{revtex4-1}

\usepackage{graphicx}
\usepackage{dcolumn}
\usepackage{bm}

\begin{document}

\preprint{NRL-TMartin-v1.4}

\title{Sonic Gradient Index Lens for Aqueous Applications}

\author{Theodore P. Martin}
 \affiliation{Acoustics Division, Naval Research Laboratory, Washington, DC 20375, USA}
\author{Michael Nicholas}
 \affiliation{Acoustics Division, Naval Research Laboratory, Washington, DC 20375, USA}
\author{Gregory J. Orris}
 \affiliation{Acoustics Division, Naval Research Laboratory, Washington, DC 20375, USA}
 \author{Liang-Wu Cai}
 \affiliation{Department of Mechanical and Nuclear Engineering, Kansas State University, Manhattan, Kansas 66506, USA}
\author{Daniel Torrent}
  \affiliation{Wave Phenomena Group, Department of Electronic Engineering, Universidad Politecnica de Valencia, C/ Camino de Vera s7n, E-46022 Valencia, Spain}
\author{Jos\'e S\'anchez-Dehesa}
  \affiliation{Wave Phenomena Group, Department of Electronic Engineering, Universidad Politecnica de Valencia, C/ Camino de Vera s7n, E-46022 Valencia, Spain}

\begin{abstract}
We study the acoustic scattering properties of a phononic crystal designed to behave as a gradient index lens in water, both experimentally and theoretically.  The gradient index lens is designed using a square lattice of stainless-steel cylinders based on a multiple scattering approach in the homogenization limit. We experimentally demonstrate that the lens follows the graded index equations derived for optics by mapping the pressure intensity generated from a spherical source at 20 kHz.  We find good agreement between the experimental result and theoretical modeling based on multiple scattering theory.   
\end{abstract}
\maketitle

Composed of ordered arrays of scatterers similar to atoms in a conventional solid, phononic crystals (PnC) are a class of metamaterial designed to control acoustic wave propagation in a medium. PnCs have been proposed for a broad range of applications in wave acoustics, with acoustic lensing~\cite{cervera02,hu05,yang04,torrent07,cai07,li09,qiu05,deng09,zhang09,lin09b,ke07} featuring prominently in the literature due in part to the ease with which focusing can be achieved by altering a crystal's gemoetric shape~\cite{cervera02,hu05} or compositional structure.~\cite{cai07,torrent07,li09} Although negative index lenses have received much attention due to their potential for near-field imaging,~\cite{qiu05,deng09,zhang09} some positive index solutions such as the acoustic analogue of the optical graded index lens~\cite{torrent07,lin09b} have not yet been explored experimentally. In addition, the majority of PnC experiments have been performed in air,~\cite{cervera02,li09,torrent06a,ke07} where the large density contrast with respect to the constituent scatterers in the crystal (typically metals) allows the scatterers to be treated as rigid.  We demonstrate below that despite the physical limitation in impedance contrasts between an aqueous medium and the scattering elements, it is possible to design a PnC that behaves as an ideal graded index lens (GIL) in water based on a fully elastic multiple scattering theory (MST).~\cite{torrent06a,torrent07,krokhin03} 

\begin{figure}
\includegraphics[width=8.6cm]{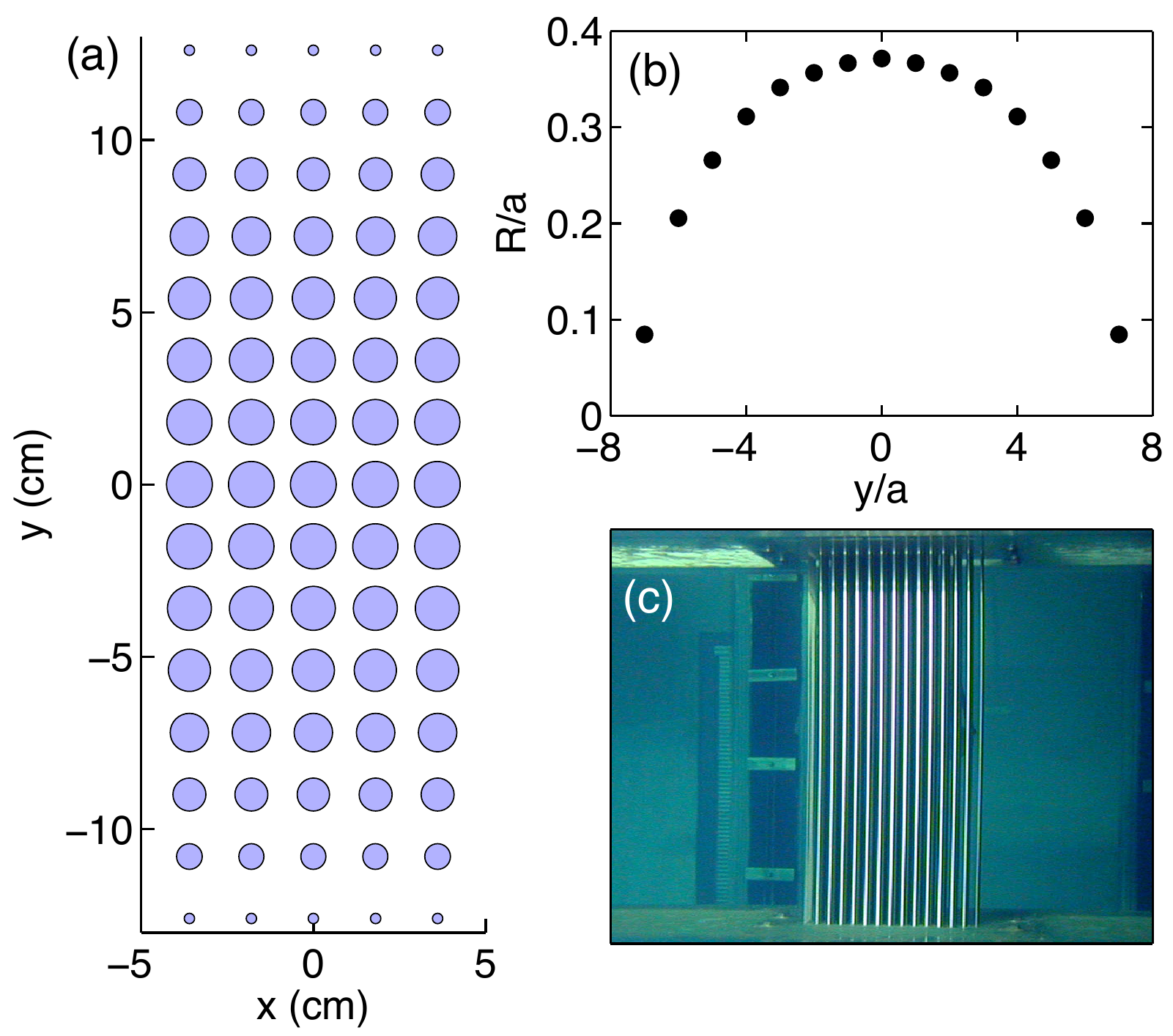}
\caption{\label{fig:1}(a) Plan schematic of the gradient index lens.  (b) Cylinder radius $R$ plotted vs position along the $y$-axis.  (c) Digital photograph of the GIL in the isolation tank.}
\end{figure}

Figure~\ref{fig:1}(a) shows a plan view schematic of the GIL design.  The axes of Fig.~\ref{fig:1}(a) and throughout the paper are oriented with the lens center at position $(x,y)=(0,0)$.  The GIL is made up of 75 stainless steel cylinders (T-316) that are 75~cm in length and arranged in a square lattice with spacing $a=1.8$~cm and dimensions $5a\times15a$.  Figure~\ref{fig:1}(b) plots the cylinder radii $R(y)$, which are stepped toward zero at each successive layer above and below the central axis ($y=0$) of the GIL.  In the homogenization limit (propagation wavelength $\lambda \gtrsim 4a$) each stratified layer in Fig.~\ref{fig:1}(a) can be treated as an effective medium.  MST~\cite{torrent07} is used to calculate each layer's effective sound speed $c_{eff}$, which is inversely proportional to the filling fraction of the cylinders.  The layers will have an effective refractive index $n_{eff}=c_{b}/c_{eff}$ ($c_{b}=1470$~m/s is the sound speed in water) that is maximal at the center of the GIL and decreases to that of water at the edges.  Our choice of $R(y)$ in Fig.~\ref{fig:1}(b) produces a graded $n_{eff}$ that obeys the same relation as an optical GIL,~\cite{smith00} $n_{eff}=n_{0}(1-\alpha^{2}y^{2})^{1/2}$, where $n_{0}$ is the refractive index at the central layer.  Our design results in $n_{0}=1.2$ and $\alpha=0.04$~cm$^{-1}$.  

Figure~\ref{fig:1}(c) shows an image of the GIL submerged in a $6\times6\times4$~m$^{3}$ isolation tank.  The cylinders are mounted between reinforced Plexiglas plates to provide stability.  Acoustic waves are produced by a 10~cm-diameter spherical source at 20~kHz ($\lambda\simeq4a$).  The wave propagation is measured in the time-domain using monitoring hydrophones at a sampling rate of 1~MHz.  Hydrophones are mounted to the source and onto a translational 3-axis Velmex VXM$^{\textregistered}$ positioning system.  The transmission intensity is measured by averaging over a 10-cycle pulse from the source; this pulse is long enough to approximate a continuous wave measurement, while being short enough to prevent contamination from reflections off the surfaces of the tank.  We have experimentally verified that the intensity $P_{0}$ produced by the source in the absence of the GIL drops radially in proportion to $1/r^{2}$.  

\begin{figure}
\includegraphics[width=8.6cm]{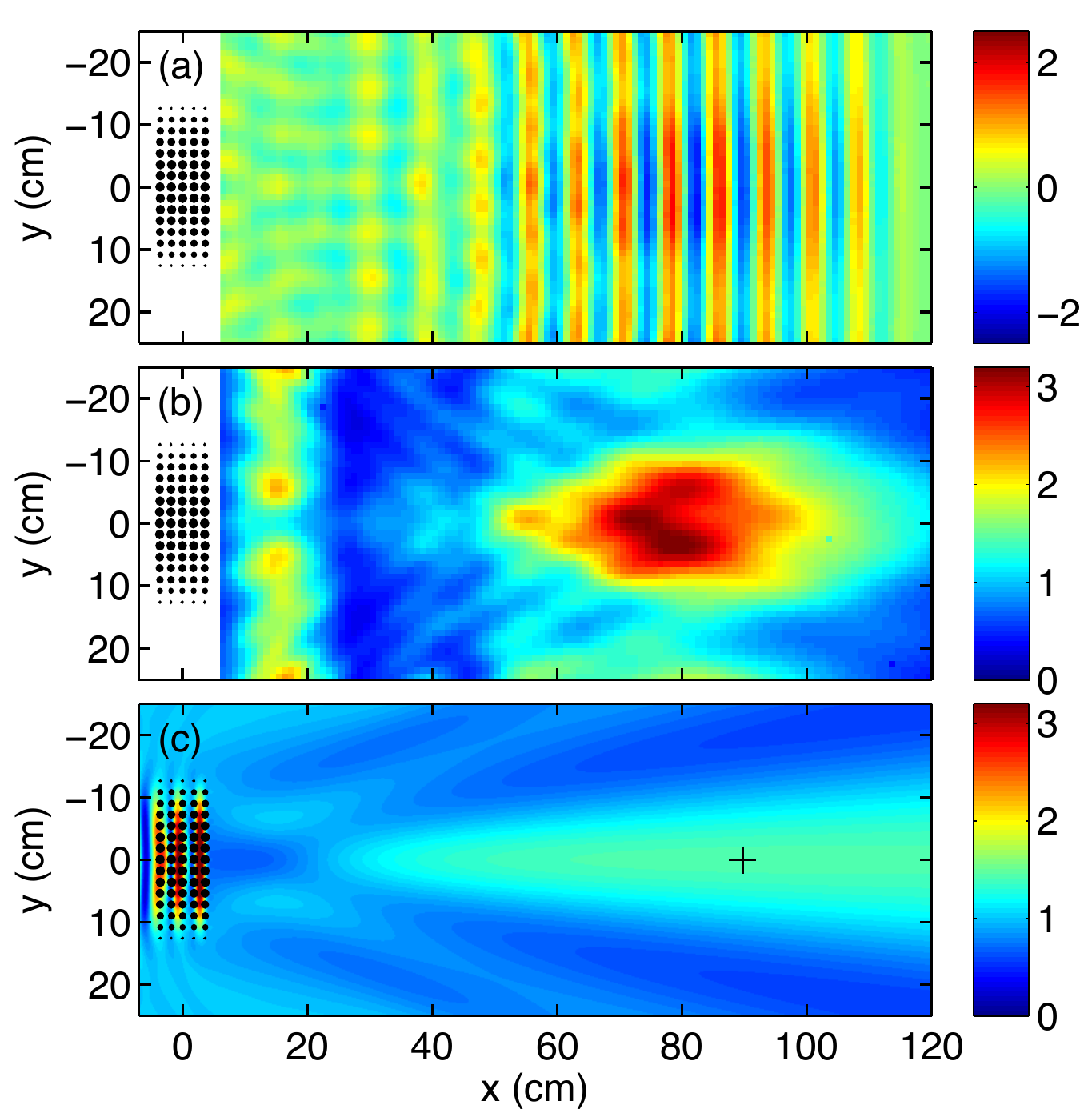}
\caption{\label{fig:2} (a) Normalized pressure amplitude $P/P_{0}$ plotted vs $x$ and $y$, measured 2.144~ms after the initial pulse leaves the source.  (b) Measured, normalized pressure intensity $(P/P_{0})^{2}$ plotted vs $x$ and $y$ after averaging over a 10-cycle pulse.  (c) Normalized pressure intensity $(P/P_{0})^{2}$ calculated using MST.  The focusing peak maximum is marked by a $+$. }
\end{figure}

Figure~\ref{fig:2}(a) shows the normalized pressure amplitude $P/P_{0}$ measured after transmission through the GIL on the side opposite the source ($x>0$).  The source is located at $(x,y)=(-196,0)$~cm, and both the source and the translational hydrophone are positioned in the plane bisecting the axial center of the cylinders.  The GIL is shown schematically to scale and at its proper location in each figure throughout the paper.  The data in Fig.~\ref{fig:2}(a) is measured 2.144~ms after the initial cycle began to leave the source.  This time gives a snapshot when the pulse is centered on an enhancement in signal amplitude observed in the vicinity of $x\simeq80$~cm.  

Figure~\ref{fig:2}(b) shows the normalized intensity averaged over the 10-cycle pulse and obtained from the same data set shown in Fig.~\ref{fig:2}(a).  As in Fig.~\ref{fig:2}(a), a clear focusing peak is observed centered close to $x\simeq80$~cm.  In Fig.~\ref{fig:2}(c) we show a two-dimensional MST calculation~\cite{torrent06a,torrent07} of the total pressure intensity (incident plus scattered) derived by placing a continuous-wave cylindrical source at $(x,y)=(-196,0)$~cm.  The calculation assumes the cylinders to be a penetrable elastic.~\cite{torrent07,krokhin03}  As with the experimental data, the simulated pressure intensity is normalized to that of the source in the absence of the GIL.  The source amplitude is a Hankel function $P_{0}=H^{(1)}_{0}(kr)$ with wavevector $k=\pi/2a$.  The MST simulation also shows a clear focusing peak, but with two important differences: (1) the measured intensity is $\sim2$ times larger than the simulation, and (2) the simulated focusing peak is slightly farther from the GIL and decays more slowly.

\begin{figure}
\includegraphics[width=8.6cm]{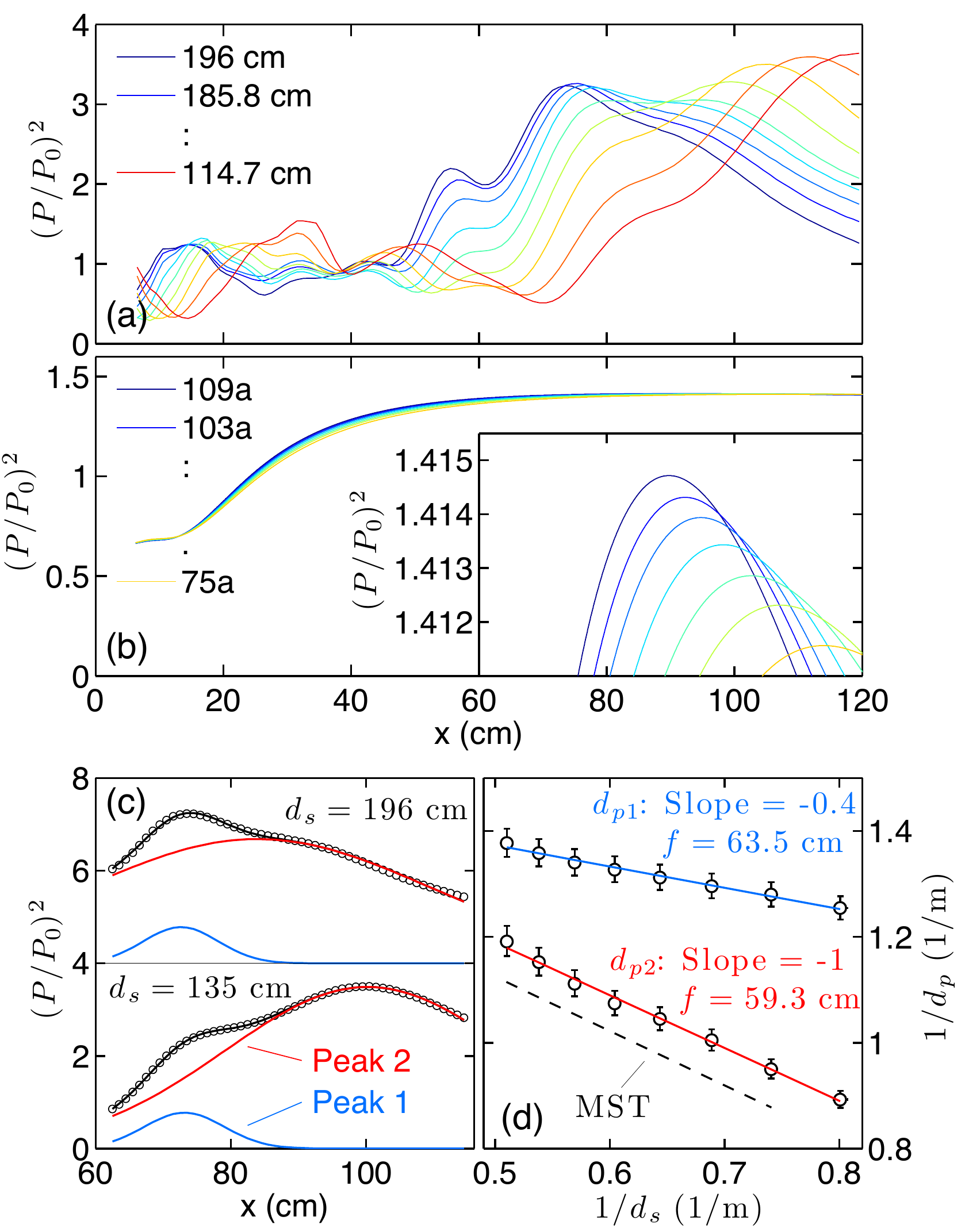}
\caption{\label{fig:3} (a) Measured, normalized pressure intensity vs $x$ for source positions $d_{s}=196$, $185.8$, $175.7$, $165.5$, $155.4$, $145.2$, $135$, $124.9$, and $114.7$~cm. (b) Normalized pressure intensity calculated using MST for source positions $d_{s}=109a$, $103a$, $98a$, $92a$, $86a$, $81a$, and $75a$.  Inset: expanded region of the $y$-axis showing the focusing peak positions.  (c) Two focusing peaks from panel (a) are replotted as circles (upper region offset for clarity).  Black lines indicate a double-gaussian fit, with blue (Peak 1) and red (Peak 2) lines showing the component gaussians individually.  (d) Inverse peak positions $1/d_{p1,2}$ plotted vs inverse source positions $1/d_{s}$.  Blue and red lines are fits to the trends of Peaks 1 and 2 respectively.  The dashed line plots the peak positions of the simulated data in panel (b). }
\end{figure}

To quantify whether our GIL design behaves as an ideal lens, in Fig.~\ref{fig:3}(a) we present measurements of the focusing peak along the central axis of the lens ($y=0$) for different source positions $d_{s}$.  For each $d_{s}$, a large-amplitude peak is observed above $x\gtrsim60$~cm, while smaller peaks are also observed closer to the GIL.  As the source is moved closer to the GIL, the large-amplitude peak moves away in qualitative agreement with the expected behavior of a lens.  In Fig.~\ref{fig:3}(b) we show MST calculations along $y=0$ for source positions similar to those in the experiment.  On initial inspection it appears that the theory shows slowly decaying focusing peaks that change very little with $d_{s}$.  However, expansion of the region around the focusing peaks [Fig.~\ref{fig:3}(b) inset] reveals that the peak positions move away in a manner similar to the experiment.

We now analyze the experimental data in Fig.~\ref{fig:3}(a) above $x>62$~cm to determine whether the focusing positions in this region follow the ideal lens equations. For an ideal lens, the focusing peak positions $d_{p}$ should scale with $d_{s}$ as $1/d_{p}=1/f-1/d_{s}$.  The focal length $f$ of a GIL can be approximated as,~\cite{smith00}
\begin{eqnarray}
\label{eqn:1} f \approx \frac{1}{n_{0} \alpha \sin \alpha t}
\end{eqnarray}
where $t=5a$ is the thickness of the GIL.  Equation~(\ref{eqn:1}) gives an estimate of $f=58.9$~cm using values of $n_{0}$ and $\alpha$ calculated in the effective medium approximation.~\footnote{Although we are technically measuring the \emph{back focal length} of the GIL, $bfl=f\cos \alpha t$, the difference between our estimate of $f$ and the sum $bfl+t/2$ is less than $1$~cm, which is less than both our experimental error and our data resolution ($=1$~cm). Thus we ignore this subtlety and compare to $f$ directly.  See Ref.~\citenum{smith00} for details.}

A close inspection of Fig.~\ref{fig:3}(a) reveals that the data above $x>62$~cm is actually composed of two superimposed peaks that both move to larger $x$ as a function of $d_{s}$.  Figure~\ref{fig:3}(c) shows two examples of a double-gaussian fit to the data in this region using a standard unconstrained, nonlinear optimization routine.  The gaussians resulting from the fit are shown individually (blue and red) in addition to the combined fit (black).  Although the fit equation $\gamma_{1}e^{-\beta_{1}(x-d_{p1})^{2}}+\gamma_{2}e^{-\beta_{2}(x-d_{p2})^{2}}$ contains six free parameters, the purpose of the fit is to obtain an estimate of the peak positions $d_{p1,2}$ and their relative amplitudes $\gamma_{1,2}$.  Fig.~\ref{fig:3}(c) demonstrates that the data is well described by the double-gaussian, with a low-amplitude peak (Peak 1) closer to the GIL and a larger-amplitude peak (Peak 2) farther away.  In both cases the amplitude of Peak 2 is about three times larger than Peak 1, suggesting that Peak 2 is the main focusing peak of the GIL.  The relative amplitudes of Peaks 1 and 2 are observed to follow the same qualitative behavior for all the source positions in Fig.~\ref{fig:3}(a).

Figure~\ref{fig:3}(d) plots the inverse positions $1/d_{p1,2}$ extracted from the gaussian fits as a function of $1/d_{s}$.  An ideal lens will produce a linear trend with a slope of $-1$ and an intercept of $1/f$.  Although the trends for both peaks are linear and have intercepts that yield similar focal lengths, the slope of Peak 1 is much less than that of an ideal lens.  However, the trend for Peak 2 results in a slope of $-1$ and its focal length $f=59.3\pm1.5$~cm agrees with the estimate of $f$ calculated using Eqn.~(\ref{eqn:1}).  The dashed line in Fig.~\ref{fig:3}(d) plots the peak locations obtained from the MST calculations in Fig.~\ref{fig:3}(b).  The theory produces a slope of $-1$ and focal length $f=61.1$~cm that closely match both the measured data and the ideal lens equations.  

We propose that Peak 1 and the other low-amplitude peaks in Fig.~\ref{fig:3}(a) are the result of constructive interference between waves scattered from the support structure of the lens.  Low-amplitude, circular interference fringes can be observed in Fig.~\ref{fig:2}(a) emanating from above and below the plot area centered at $x\simeq25$~cm.  These fringes are the result of scattering off of stabilizing pillars at the corners of the GIL support structure.  While averaging over a few initial pulse cycles will reduce the interference, a small number of cycles gives a poor approximation to a continuous wave measurement and limits the number of multiple scattering events that contribute to the focusing peak.  Therefore we have chosen to average over many cycles and rely on the gaussian fitting routine to remove the spurious interference.   

\begin{figure}
\includegraphics[width=8.6cm]{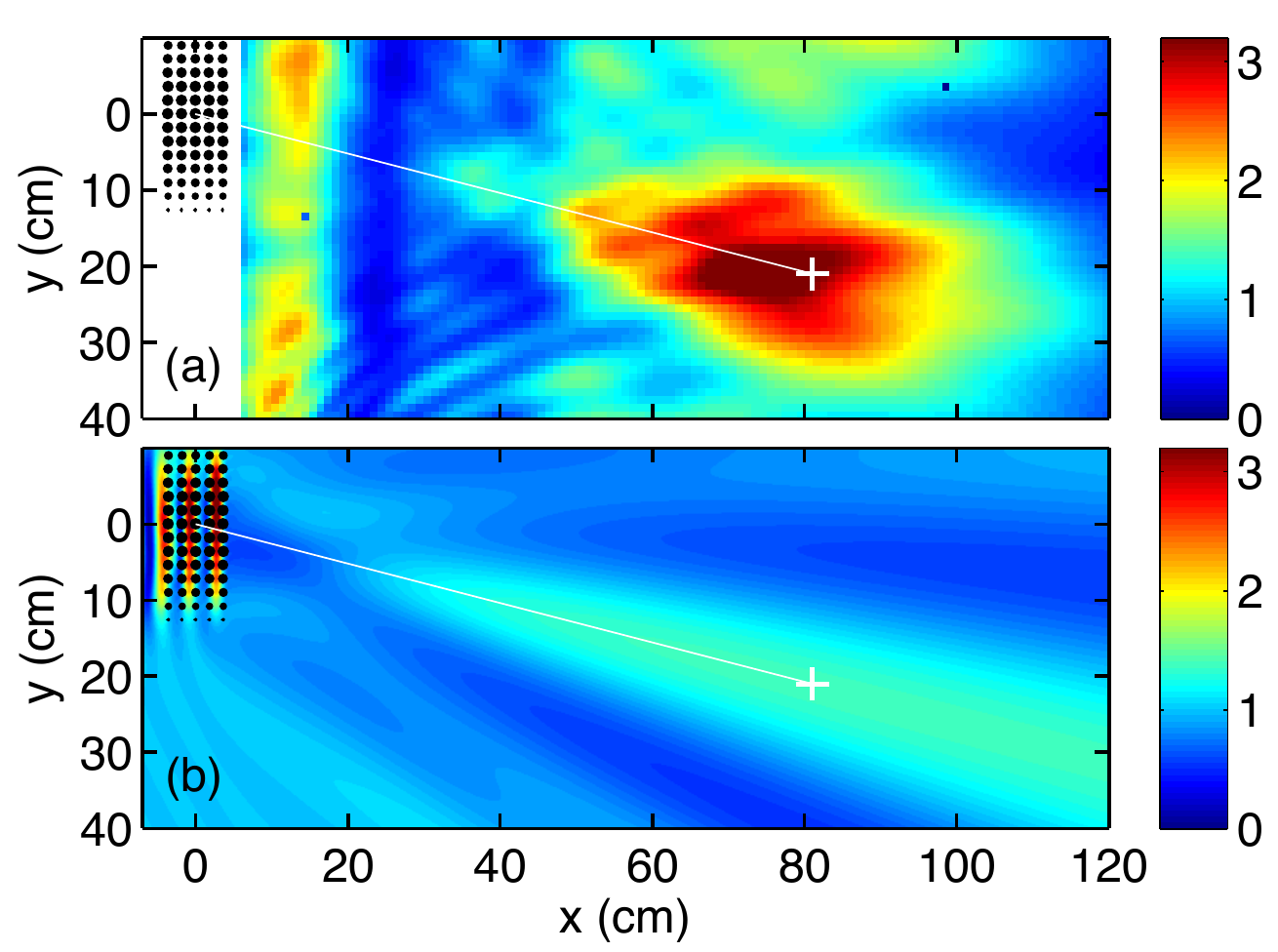}
\caption{\label{fig:4} (a) Measured, normalized pressure intensity $(P/P_{0})^{2}$ plotted vs $x$ and $y$ with the source at a $14.7\,^{\circ}$ angle with respect to the $x$-axis.  (b) Normalized pressure intensity $(P/P_{0})^{2}$ calculated using multiple scattering theory with the source at a $15\,^{\circ}$ angle.  White lines indicate the off-axis angle; positions of the expected focusing peaks are marked with a $+$. }
\end{figure}

Figure~\ref{fig:4} demonstrates that our GIL design acts as a lens with the source off the central axis.  Figure~\ref{fig:4}(a) shows $(P/P_{0})^{2}$ measured with the spherical source located at a $14.7\,^{\circ}$ angle with respect to the origin.  Figure~\ref{fig:4}(b) shows the MST calculation for the same source location.  Thin white lines in Figs.~\ref{fig:4}(a,b) indicate a $14.7\,^{\circ}$ angle with respect to the $x$-axis and extend to the expected focusing positions of an ideal lens with $f\approx60$~cm.  Both the measured data and the MST calculation demonstrate a strong focusing peak at the expected location.  Interference fringes from the support pillar can also be observed superimposed on the focusing peak in Fig.~\ref{fig:4}(a).

In summary, we have designed and constructed a gradient index lens that operates in water at sonic frequencies.  Our transmission measurements demonstrate that our GIL design focuses as an ideal lens based on the optical GIL equations.  Our measurements are also consistent with the focusing positions obtained from two-dimensional models using multiple scattering theory.  We emphasize that our GIL behaves as an ideal lens at the limit of homogenization ($\lambda\simeq4a$) and with a thickness on the order of a wavelength ($t=5\lambda/4$).  Such performance at the limit of homogenization theory demonstrates the versatility of phononic crystals designed using multiple scattering theory.

This work was supported by the U.S. Office of Naval Research.

\clearpage

\bibliography{TMartin_bib}

\begin{thebibliography}{15}%
\makeatletter
\providecommand \@ifxundefined [1]{%
 \@ifx{#1\undefined}
}%
\providecommand \@ifnum [1]{%
 \ifnum #1\expandafter \@firstoftwo
 \else \expandafter \@secondoftwo
 \fi
}%
\providecommand \@ifx [1]{%
 \ifx #1\expandafter \@firstoftwo
 \else \expandafter \@secondoftwo
 \fi
}%
\providecommand \natexlab [1]{#1}%
\providecommand \enquote  [1]{``#1''}%
\providecommand \bibnamefont  [1]{#1}%
\providecommand \bibfnamefont [1]{#1}%
\providecommand \citenamefont [1]{#1}%
\providecommand \href@noop [0]{\@secondoftwo}%
\providecommand \href [0]{\begingroup \@sanitize@url \@href}%
\providecommand \@href[1]{\@@startlink{#1}\@@href}%
\providecommand \@@href[1]{\endgroup#1\@@endlink}%
\providecommand \@sanitize@url [0]{\catcode `\\12\catcode `\$12\catcode
  `\&12\catcode `\#12\catcode `\^12\catcode `\_12\catcode `\%12\relax}%
\providecommand \@@startlink[1]{}%
\providecommand \@@endlink[0]{}%
\providecommand \url  [0]{\begingroup\@sanitize@url \@url }%
\providecommand \@url [1]{\endgroup\@href {#1}{\urlprefix }}%
\providecommand \urlprefix  [0]{URL }%
\providecommand \Eprint [0]{\href }%
\@ifxundefined \urlstyle {%
  \providecommand \doi  [0]{\begingroup \@sanitize@url \@doi}%
  \providecommand \@doi [1]{\endgroup \@@startlink {\doibase
  #1}doi:\discretionary {}{}{}#1\@@endlink }%
}{%
  \providecommand \doi  [0]{doi:\discretionary{}{}{}\begingroup
  \urlstyle{rm}\Url }%
}%
\providecommand \doibase [0]{http://dx.doi.org/}%
\providecommand \Doi [0]{\begingroup \@sanitize@url \@Doi }%
\providecommand \@Doi  [1]{\endgroup\@@startlink{\doibase#1}\@@Doi}%
\providecommand \@@Doi [1]{#1\@@endlink}%
\providecommand \selectlanguage [0]{\@gobble}%
\providecommand \bibinfo  [0]{\@secondoftwo}%
\providecommand \bibfield  [0]{\@secondoftwo}%
\providecommand \translation [1]{[#1]}%
\providecommand \BibitemOpen [0]{}%
\providecommand \bibitemStop [0]{}%
\providecommand \bibitemNoStop [0]{.\EOS\space}%
\providecommand \EOS [0]{\spacefactor3000\relax}%
\providecommand \BibitemShut  [1]{\csname bibitem#1\endcsname}%
\bibitem [{\citenamefont {Cervera}\ \emph {et~al.}(2002)\citenamefont
  {Cervera}, \citenamefont {Sanchis}, \citenamefont {S\'anchez-P\'erez},
  \citenamefont {Mart\'inez-Sala}, \citenamefont {Rubio}, \citenamefont
  {Meseguer}, \citenamefont {L\'opez}, \citenamefont {Caballero},\ and\
  \citenamefont {S\'anchez-Dehesa}}]{cervera02}%
  \BibitemOpen
  \bibfield  {author} {\bibinfo {author} {\bibfnamefont {F.}~\bibnamefont
  {Cervera}}, \bibinfo {author} {\bibfnamefont {L.}~\bibnamefont {Sanchis}},
  \bibinfo {author} {\bibfnamefont {J.~V.}\ \bibnamefont {S\'anchez-P\'erez}},
  \bibinfo {author} {\bibfnamefont {R.}~\bibnamefont {Mart\'inez-Sala}},
  \bibinfo {author} {\bibfnamefont {C.}~\bibnamefont {Rubio}}, \bibinfo
  {author} {\bibfnamefont {F.}~\bibnamefont {Meseguer}}, \bibinfo {author}
  {\bibfnamefont {C.}~\bibnamefont {L\'opez}}, \bibinfo {author} {\bibfnamefont
  {D.}~\bibnamefont {Caballero}}, \ and\ \bibinfo {author} {\bibfnamefont
  {J.}~\bibnamefont {S\'anchez-Dehesa}},\ }\href@noop {} {\bibfield  {journal}
  {\bibinfo  {journal} {Phys. Rev. Lett.},\ }\textbf {\bibinfo {volume} {88}},\
  \bibinfo {pages} {023902} (\bibinfo {year} {2002})}\BibitemShut {NoStop}%
\bibitem [{\citenamefont {Hu}\ and\ \citenamefont {Chan}(2005)}]{hu05}%
  \BibitemOpen
  \bibfield  {author} {\bibinfo {author} {\bibfnamefont {X.}~\bibnamefont
  {Hu}}\ and\ \bibinfo {author} {\bibfnamefont {C.~T.}\ \bibnamefont {Chan}},\
  }\href@noop {} {\bibfield  {journal} {\bibinfo  {journal} {Phys. Rev.
  Lett.},\ }\textbf {\bibinfo {volume} {95}},\ \bibinfo {pages} {154501}
  (\bibinfo {year} {2005})}\BibitemShut {NoStop}%
\bibitem [{\citenamefont {Yang}\ \emph {et~al.}(2004)\citenamefont {Yang},
  \citenamefont {Page}, \citenamefont {Liu}, \citenamefont {Cowan},
  \citenamefont {Chan},\ and\ \citenamefont {Sheng}}]{yang04}%
  \BibitemOpen
  \bibfield  {author} {\bibinfo {author} {\bibfnamefont {S.}~\bibnamefont
  {Yang}}, \bibinfo {author} {\bibfnamefont {J.~H.}\ \bibnamefont {Page}},
  \bibinfo {author} {\bibfnamefont {Z.}~\bibnamefont {Liu}}, \bibinfo {author}
  {\bibfnamefont {M.~L.}\ \bibnamefont {Cowan}}, \bibinfo {author}
  {\bibfnamefont {C.~T.}\ \bibnamefont {Chan}}, \ and\ \bibinfo {author}
  {\bibfnamefont {P.}~\bibnamefont {Sheng}},\ }\href@noop {} {\bibfield
  {journal} {\bibinfo  {journal} {Phys. Rev. Lett.},\ }\textbf {\bibinfo
  {volume} {93}},\ \bibinfo {pages} {024301} (\bibinfo {year}
  {2004})}\BibitemShut {NoStop}%
\bibitem [{\citenamefont {Torrent}\ and\ \citenamefont
  {S\'anchez-Dehesa}(2007)}]{torrent07}%
  \BibitemOpen
  \bibfield  {author} {\bibinfo {author} {\bibfnamefont {D.}~\bibnamefont
  {Torrent}}\ and\ \bibinfo {author} {\bibfnamefont {J.}~\bibnamefont
  {S\'anchez-Dehesa}},\ }\href@noop {} {\bibfield  {journal} {\bibinfo
  {journal} {New J. Phys.},\ }\textbf {\bibinfo {volume} {9}},\ \bibinfo
  {pages} {323} (\bibinfo {year} {2007})}\BibitemShut {NoStop}%
\bibitem [{\citenamefont {Cai}\ \emph {et~al.}(2007)\citenamefont {Cai},
  \citenamefont {Liu}, \citenamefont {He},\ and\ \citenamefont {Liu}}]{cai07}%
  \BibitemOpen
  \bibfield  {author} {\bibinfo {author} {\bibfnamefont {F.}~\bibnamefont
  {Cai}}, \bibinfo {author} {\bibfnamefont {F.}~\bibnamefont {Liu}}, \bibinfo
  {author} {\bibfnamefont {Z.}~\bibnamefont {He}}, \ and\ \bibinfo {author}
  {\bibfnamefont {Z.}~\bibnamefont {Liu}},\ }\href@noop {} {\bibfield
  {journal} {\bibinfo  {journal} {Appl. Phys. Lett.},\ }\textbf {\bibinfo
  {volume} {91}},\ \bibinfo {pages} {203515} (\bibinfo {year}
  {2007})}\BibitemShut {NoStop}%
\bibitem [{\citenamefont {Li}\ \emph {et~al.}(2009)\citenamefont {Li},
  \citenamefont {Fok}, \citenamefont {Yin}, \citenamefont {Bartal},\ and\
  \citenamefont {Zhang}}]{li09}%
  \BibitemOpen
  \bibfield  {author} {\bibinfo {author} {\bibfnamefont {J.}~\bibnamefont
  {Li}}, \bibinfo {author} {\bibfnamefont {L.}~\bibnamefont {Fok}}, \bibinfo
  {author} {\bibfnamefont {X.}~\bibnamefont {Yin}}, \bibinfo {author}
  {\bibfnamefont {G.}~\bibnamefont {Bartal}}, \ and\ \bibinfo {author}
  {\bibfnamefont {X.}~\bibnamefont {Zhang}},\ }\href@noop {} {\bibfield
  {journal} {\bibinfo  {journal} {Nature Mater.},\ }\textbf {\bibinfo {volume}
  {8}},\ \bibinfo {pages} {931} (\bibinfo {year} {2009})}\BibitemShut {NoStop}%
\bibitem [{\citenamefont {Qiu}\ \emph {et~al.}(2005)\citenamefont {Qiu},
  \citenamefont {Zhang},\ and\ \citenamefont {Liu}}]{qiu05}%
  \BibitemOpen
  \bibfield  {author} {\bibinfo {author} {\bibfnamefont {C.}~\bibnamefont
  {Qiu}}, \bibinfo {author} {\bibfnamefont {X.}~\bibnamefont {Zhang}}, \ and\
  \bibinfo {author} {\bibfnamefont {Z.}~\bibnamefont {Liu}},\ }\href@noop {}
  {\bibfield  {journal} {\bibinfo  {journal} {Phys. Rev. B},\ }\textbf
  {\bibinfo {volume} {71}},\ \bibinfo {pages} {054302} (\bibinfo {year}
  {2005})}\BibitemShut {NoStop}%
\bibitem [{\citenamefont {Deng}\ \emph {et~al.}(2009)\citenamefont {Deng},
  \citenamefont {Ding}, \citenamefont {He}, \citenamefont {Zhao}, \citenamefont
  {Shi},\ and\ \citenamefont {Liu}}]{deng09}%
  \BibitemOpen
  \bibfield  {author} {\bibinfo {author} {\bibfnamefont {K.}~\bibnamefont
  {Deng}}, \bibinfo {author} {\bibfnamefont {Y.}~\bibnamefont {Ding}}, \bibinfo
  {author} {\bibfnamefont {Z.}~\bibnamefont {He}}, \bibinfo {author}
  {\bibfnamefont {H.}~\bibnamefont {Zhao}}, \bibinfo {author} {\bibfnamefont
  {J.}~\bibnamefont {Shi}}, \ and\ \bibinfo {author} {\bibfnamefont
  {Z.}~\bibnamefont {Liu}},\ }\href@noop {} {\bibfield  {journal} {\bibinfo
  {journal} {J. Phys. D: Appl. Phys.},\ }\textbf {\bibinfo {volume} {42}},\
  \bibinfo {pages} {185505} (\bibinfo {year} {2009})}\BibitemShut {NoStop}%
\bibitem [{\citenamefont {Zhang}\ \emph {et~al.}(2009)\citenamefont {Zhang},
  \citenamefont {Yin},\ and\ \citenamefont {Fang}}]{zhang09}%
  \BibitemOpen
  \bibfield  {author} {\bibinfo {author} {\bibfnamefont {S.}~\bibnamefont
  {Zhang}}, \bibinfo {author} {\bibfnamefont {L.}~\bibnamefont {Yin}}, \ and\
  \bibinfo {author} {\bibfnamefont {N.}~\bibnamefont {Fang}},\ }\href@noop {}
  {\bibfield  {journal} {\bibinfo  {journal} {Phys. Rev. Lett.},\ }\textbf
  {\bibinfo {volume} {102}},\ \bibinfo {pages} {194301} (\bibinfo {year}
  {2009})}\BibitemShut {NoStop}%
\bibitem [{\citenamefont {Lin}\ \emph {et~al.}(2009)\citenamefont {Lin},
  \citenamefont {Huang}, \citenamefont {Sun},\ and\ \citenamefont
  {Wu}}]{lin09b}%
  \BibitemOpen
  \bibfield  {author} {\bibinfo {author} {\bibfnamefont {S.-C.~S.}\
  \bibnamefont {Lin}}, \bibinfo {author} {\bibfnamefont {T.~J.}\ \bibnamefont
  {Huang}}, \bibinfo {author} {\bibfnamefont {J.-H.}\ \bibnamefont {Sun}}, \
  and\ \bibinfo {author} {\bibfnamefont {T.-T.}\ \bibnamefont {Wu}},\
  }\href@noop {} {\bibfield  {journal} {\bibinfo  {journal} {Phys. Rev. B},\
  }\textbf {\bibinfo {volume} {79}},\ \bibinfo {pages} {094302} (\bibinfo
  {year} {2009})}\BibitemShut {NoStop}%
\bibitem [{\citenamefont {Ke}\ \emph {et~al.}(2007)\citenamefont {Ke},
  \citenamefont {Liu}, \citenamefont {Pang}, \citenamefont {Qiu}, \citenamefont
  {Zhao}, \citenamefont {Peng},\ and\ \citenamefont {Shi}}]{ke07}%
  \BibitemOpen
  \bibfield  {author} {\bibinfo {author} {\bibfnamefont {M.}~\bibnamefont
  {Ke}}, \bibinfo {author} {\bibfnamefont {Z.}~\bibnamefont {Liu}}, \bibinfo
  {author} {\bibfnamefont {P.}~\bibnamefont {Pang}}, \bibinfo {author}
  {\bibfnamefont {C.}~\bibnamefont {Qiu}}, \bibinfo {author} {\bibfnamefont
  {D.}~\bibnamefont {Zhao}}, \bibinfo {author} {\bibfnamefont {S.}~\bibnamefont
  {Peng}}, \ and\ \bibinfo {author} {\bibfnamefont {J.}~\bibnamefont {Shi}},\
  }\href@noop {} {\bibfield  {journal} {\bibinfo  {journal} {Appl. Phys.
  Lett.},\ }\textbf {\bibinfo {volume} {90}},\ \bibinfo {pages} {083509}
  (\bibinfo {year} {2007})}\BibitemShut {NoStop}%
\bibitem [{\citenamefont {Torrent}\ \emph {et~al.}(2006)\citenamefont
  {Torrent}, \citenamefont {Hakansson}, \citenamefont {Cervera},\ and\
  \citenamefont {S\'anchez-Dehesa}}]{torrent06a}%
  \BibitemOpen
  \bibfield  {author} {\bibinfo {author} {\bibfnamefont {D.}~\bibnamefont
  {Torrent}}, \bibinfo {author} {\bibfnamefont {A.}~\bibnamefont {Hakansson}},
  \bibinfo {author} {\bibfnamefont {F.}~\bibnamefont {Cervera}}, \ and\
  \bibinfo {author} {\bibfnamefont {J.}~\bibnamefont {S\'anchez-Dehesa}},\
  }\href@noop {} {\bibfield  {journal} {\bibinfo  {journal} {Phys. Rev.
  Lett.},\ }\textbf {\bibinfo {volume} {96}},\ \bibinfo {pages} {204302}
  (\bibinfo {year} {2006})}\BibitemShut {NoStop}%
\bibitem [{\citenamefont {Krokhin}\ \emph {et~al.}(2003)\citenamefont
  {Krokhin}, \citenamefont {Arriaga},\ and\ \citenamefont {Gumen}}]{krokhin03}%
  \BibitemOpen
  \bibfield  {author} {\bibinfo {author} {\bibfnamefont {A.~A.}\ \bibnamefont
  {Krokhin}}, \bibinfo {author} {\bibfnamefont {J.}~\bibnamefont {Arriaga}}, \
  and\ \bibinfo {author} {\bibfnamefont {L.~N.}\ \bibnamefont {Gumen}},\
  }\href@noop {} {\bibfield  {journal} {\bibinfo  {journal} {Phys. Rev.
  Lett.},\ }\textbf {\bibinfo {volume} {91}},\ \bibinfo {pages} {264302}
  (\bibinfo {year} {2003})}\BibitemShut {NoStop}%
\bibitem [{\citenamefont {Smith}(2000)}]{smith00}%
  \BibitemOpen
  \bibfield  {author} {\bibinfo {author} {\bibfnamefont {W.~J.}\ \bibnamefont
  {Smith}},\ }\href@noop {} {\emph {\bibinfo {title} {Modern Optical
  Engineering}}}\ (\bibinfo  {publisher} {McGraw-Hill, Inc.},\ \bibinfo {year}
  {2000})\BibitemShut {NoStop}%
\bibitem [{Note1()}]{Note1}%
  \BibitemOpen
  \bibinfo {note} {Although we are technically measuring the \protect \emph
  {back focal length} of the GIL, $bfl=f\protect \qopname \relax o{cos}\alpha
  t$, the difference between our estimate of $f$ and the sum $bfl+t/2$ is less
  than $1$~cm, which is less than both our experimental error and our data
  resolution ($=1$~cm). Thus we ignore this subtlety and compare to $f$
  directly. See Ref.~\protect \citenum {smith00} for details.}\BibitemShut
  {Stop}%
\end{thebibliography}%

\clearpage

\clearpage

\clearpage

\clearpage

\end{document}